\documentclass[pre, fleqn,twocolumn,showpacs,preprintnumbers,amsmath,amssymb]{revtex4}

\usepackage{graphicx}
\usepackage{dcolumn}
\usepackage{bm}
\usepackage{color}
\usepackage{amsmath}

\begin{document}


\title{Early stage of integrable turbulence in 1D NLS equation: the semi-classical approach to statistics}

\author{Giacomo Roberti$^1$, Gennady El$^1$, St\'ephane Randoux$^{2,3}$ and Pierre Suret$^{2,3}$}

\email[Corresponding author : ]{Pierre.Suret@univ-lille.fr}

\affiliation{$^1$ Department of Mathematics, Physics and Electrical Engineering, Northumbria  University, Newcastle upon Tyne, NE1 8ST, United Kingdom}

\affiliation{$^2$ University of Lille, CNRS, UMR 8523 -  Physique des Lasers Atomes et Mol\'ecules (PHLAM), F-59000 Lille, France}

\affiliation{$^3$ Centre d'Etudes et de Recherches Lasers et Applications (CERLA), Cit\'e scientifique, 59655 Villeneuve d'ascq Cedex, France}


\date{\today}

\begin{abstract}
We examine statistical properties of integrable turbulence in the defocusing and focusing regimes of one-dimensional small-dispersion nonlinear Schr\"odinger equation (1D-NLSE). Specifically, we study the 1D-NLSE evolution of partially coherent waves having Gaussian statistics at time $t=0$. Using  short time asymptotic expansions and taking advantage of the scale separation in the semi-classical regime we  obtain a simple explicit formula describing an early stage of the evolution of the fourth moment of the random wave field amplitude, a quantitative measure of the ``tailedness'' of the probability density function. Our results show excellent agreement with numerical simulations of the full 1D-NLSE random field dynamics and provide insight into the emergence of the well-known phenomenon of heavy (resp. low) tails of the statistical distribution emerging in the focusing (resp. defocusing) regime of 1D-NLSE. 
\end{abstract}

\maketitle

\section{Introduction}

Predicting the statistical properties of nonlinear random waves is at the core of research on turbulence~\cite{Nazarenko}. Generally speaking, nonlinear wave propagation in media with self-focusing nonlinearity tends
to produce heavy-tailed deviations from the initial Gaussian statistics,  observed in the probability density function (PDF) of the random wave field amplitude.  In recent years, the question of the emergence of heavy-tailed statistical distributions has been extensively studied in relation to the occurrence of extreme events such as  Rogue Waves (RW),  mainly in the physical contexts of fluid dynamics  \cite{kharif2009rogue, Onorato:04, Onorato:05, Onorato:13} and  optics \cite{Solli:07, Mussot:09, Walczak:15, Pierangeli:15,Suret:16, Narhi:16, Safari:17, Tikan:18}. Even though statistical
properties of nonlinear defocusing media have been less extensively examined, several experiments have shown that defocusing nonlinearities tend to produce low-tailed deviations from 
the initial Gaussian statistical distribution \cite{Randoux:14,Bromberg:10}.

Among nonlinear random waves systems, those that are described by the  one-dimensional nonlinear Schr\"odinger Equation (1D-NLSE) play a unique and fundamental role. First of all, 1D-NLSE describes various physical systems at leading order in the amplitude and secondly, 1D-NLSE is integrable and exhibit many remarkable exact solutions, including solitons (both for the focusing and the defocusing cases)~\cite{Yang}. The salient features of focusing regime of 1D-NLSE are  modulational instability of plane waves and the existence of breather solutions localized both in space and in time and thus the focusing 1D-NLSE  plays a central role in the study of RWs ~\cite{Peregrine:83, Henderson:99,Akhmediev:09,Akhmediev:09b,Kibler:10,Chabchoub:11, Kibler:11}. The study of the statistics of nonlinear wave systems described by 1D-NLSE  enters within the framework of {\it integrable turbulence} proposed by Zakharov~\cite{Zakharov:09, Zakharov:13, Agafontsev:15}. Generally, integrable turbulence arises in nonlinear random wave systems described  by integrable equations such as the 1D-NLSE, the Korteweg de Vries (KdV) equation and other physically important equations. Given the absence of resonances in integrable systems, the mechanisms underlying integrable turbulence~\cite{Walczak:15, Agafontsev:15, Suret:16, Randoux:16, SotoCrespo:16, AkhmedievIntegrableTurbulence:16, randoux2016livre, Suret:17, Tikan:18} are of profoundly different nature than those found in the standard {\it wave turbulence}~\cite{Zakharov,Nazarenko, Picozzi:14}. Since many nonlinear wave systems can be described by partial differential equations having an integrable core part, integrable turbulence has become an active field of research both from the theoretical and experimental perspective ~\cite{Zakharov:09, Zakharov:13, Agafontsev:15,Walczak:15,  Suret:16, Randoux:16, SotoCrespo:16, AkhmedievIntegrableTurbulence:16, randoux2016livre, Suret:17, Tikan:18}.

Random wave fields composed of a linear superposition of a large number of {\it independent} Fourier modes having random phases and/or amplitudes play  a fundamental role in wave turbulence~\cite{Zakharov, Nazarenko} and in integrable turbulence  ~\cite{Randoux:16, randoux2016livre}. As an immediate consequence of the  central limit theorem, these nonlinear random waves are characterized by Gaussian statistics. Numerical simulations and optical fiber experiments show that the long-term evolution  of such Gaussian random wave fields in the focusing regime of 1D-NLSE is accompanied by the emergence of a heavy-tailed PDF ~\cite{Walczak:15,Suret:16,Tikan:18}. Contrastingly, in the defocusing regime of 1D-NLSE, the long-term statistics of  random waves with the same initial distributions, is characterized by a low-tailed PDF ~\cite{Randoux:14,Randoux:16, randoux2016livre}.  The theoretical description  of statistical changes that occur in the course of  nonlinear propagation of such wave fields in 1D-NLSE systems is a very active current field of research ~\cite{Randoux:16, randoux2016livre}. 

The first statistical theory of wave systems described by the integrable 
1D-NLSE has been established within the framework of wave turbulence
theory \cite{Janssen:03}. The wave turbulence theory fundamentally relies on a scale separation where
the linear dispersion time $T_L$ is supposed to be much shorter than the nonlinear    
time $T_{NL}$. Wave turbulence theory is principally applicable
to sets of random dispersive waves that are engaged in a {\it weakly } nonlinear {\it resonant}  interaction~\cite{Nazarenko, Zakharov} (we note, howevere, that resonant energy exchanges are not possible in wave systems ruled by the integrable 1D-NLSE \cite{Janssen:03,Suret:11,Zakharov:09}). 
A modified treatment of the wave turbulence theory based on the derivation
of quasi-kinetic equations for the lowest order moments of the wave field
has been proposed in ref. \cite{Janssen:03, Soh:10,Suret:11,Picozzi:14}.  By using this theoretical approach, the evolution of the kurtosis (the fourth standardized moment of the PDF) of partially coherent waves has been derived both in the focusing and defocusing regime of the 1D-NLSE~\cite{Janssen:03}. The kurtosis quantifies the ``heaviness'' of the tail in the PDF of the field amplitude distribution and this theory confirmed that the kurtosis increases in the focusing case and decreases in the defocusing case while the initial field is characterized by Gaussian statistics. However, we stress that this kind of  analytical treatment  is inherently {\it limited to the weakly nonlinear}  propagation regime.\\

The question of integrable turbulence (i.e. statistical theory of random nonlinear
wave systems governed by integrable partial differential equations) has been recently
approached from a completely different perspective. For a certain class of wave systems, one can take advantage of the mathematical framework of dispersive
hydrodynamics -- the semi-classical theory of nonlinear dispersive waves \cite{Biondini:16} --
in order to analyse the wave evolution asymptotically, by introducing two distinct spatio-temporal
scales: the long scale specified by initial conditions and the short scale by the internal coherence
length (i.e. the typical size of the coherent soliton-like structures).

The semi-classical, dispersive hydrodynamic approach describes the propagation regimes of a completely opposite nature compared to  the
regimes considered in the framework of wave turbulence theory.
 This approach can be applied  to the 1D-NLSE propagation if the initial scale of the fluctuations of the power of the complex field $|\psi|^2$ are much larger than the one corresponding to the balance between nonlinearity and  dispersion. In  most of the standard cases, this separation of scales correspond to situations where the nonlinear part of the energy is much greater that the linear (kinetic) part of the energy at the initial time.  As shown in ref. \cite{Randoux:17}, this scale separation permits one to split the development of integrable turbulence
into two distinct stages characterized by qualitatively different dynamical and statistical features.
At the initial (we shall call it ``pre-breaking'') stage of the evolution nonlinear effects dominate linear
dispersion and the wave fronts of the random initial field
experience gradual steepening leading to the formation
of gradient catastrophes that are subsequently regularized through the
generation of dispersive shock waves in the defocusing regime \cite{El:2016} and of Peregrine-like
breather sequences in the focusing regime \cite{Tikan:17}. As shown in ref. \cite{Randoux:17}, the dynamical and
statistical features that occur at the pre-breaking stage of the defocusing 1D-NLSE can be interpreted in
terms of the evolution of random Riemann waves.\\

In this paper, we extend the analysis of the previous works based on the semi-classical
treatment of the 1D-NLSE with random initial data  by calculating the short-time evolution of the 
the normalized fourth moment $\kappa_4$ of the amplitude of the field. Similar to the standard kurtosis, the quantity $\kappa_4$ describes the degree of the deviation from the initial statistical distribution which is often assumed to be Gaussian~\cite{Onorato:16}.
Using the semi-classical Madelung transform and performing the zero dispersion limit, we derive a general  analytical expression for the short-time evolution
of  the fourth moment of the random 1D-NLSE wave field in terms of hydrodynamic
variables, and  show that this expression can be further simplified 
for the wave field having  Gaussian statistics at initial time.  Our analytical asymptotic results are shown to be in excellent agreement with numerical simulations of the evolution of partially coherent initial data in 1D-NLSE.

This paper is organised as follows :

- Sec. II : using the semi-classical approximation, we identify the initial stage of the 1D-NLSE development of partially coherent waves with the nonlinearity dominated, dispersionless regime, and derive the general expression for the  short-time evolution of the fourth order moment $\kappa_4$ as a power series expansion in time $t$;

- Sec III : we apply the derived formula for $\kappa_4$ to the fundamental case of random waves characterized by Gaussian statistics at time $t=0$;

- Sec IV : we provide a comparison between our semi-classical  analytical results and numerical simulations of 1D-NLSE.

\section{The dispersionless limit of the 1D-NLSE and the time evolution of the fourth-order moment of a random wave field}\label{sec:main}

We consider the  1D-NLSE in the normalized form
 \begin{equation}
 i\varepsilon\frac{\partial\psi}{\partial t}+\frac{\varepsilon^2}{2}\frac{\partial^2\psi}{\partial x^2}+\sigma |\psi|^2\psi=0,
 \label{eq:NLS}
 \end{equation}
 where $\psi$ is a complex field,  $\varepsilon$ is the dispersion parameter, $\sigma=-1$ in the defocusing regime and  $\sigma=+1$ in the focusing regime.

The 1D-NLSE  \eqref{eq:NLS} is considered in a periodic box of size $L$,
 $\psi(x+L,t)=\psi(x,t) \, \forall t $. The field $\psi$ then can  be
represented as a Fourier series:
\begin{equation}\label{eq:psi}
\psi(x,t)=\sum_k\psi_k(t)e^{\frac{2i\pi}{L}kx} \quad \hbox{with} \quad k\in \mathbb{Z},
\end{equation}
where the Fourier coefficients are given by 
\begin{equation}
\psi_k(t)=\frac{1}{L}\int_0^L\psi(x,t)e^{-2i\pi kx} dx.
\end{equation}

The ``density of particles'' $N$ and the
momentum $P$ represent  integrals of motion and are expressed in terms of Fourier coefficients:
\begin{equation}
 N=\frac{1}{L}\int_{0}^{L}|\psi|^2dx=\sum_k|\psi_k|^2 ,
\label{Npart} 
\end{equation}
\begin{equation}
P=\frac{1}{L}\int_{0}^{L}\psi_x\psi^*dx=\sum_k \Big(\frac{2\pi ik}{L}\Big) |\psi_k|^2 .
\label{Momentum}
\end{equation}

The Hamiltonian, that we represent in the form
\begin{equation}
  H=\varepsilon^2H_{\rm L}+ H_{\rm NL}
  \label{eq:H}
\end{equation}
is also integral of motion, which is naturally split into two parts: the linear (kinetic energy) part 
\begin{equation}
\varepsilon^2 H_{\rm L}(t)=\frac{\varepsilon^2}{2L}\int_{0}^{L}|\psi_x|^2dx=\frac{\varepsilon^2}{2}\sum_k \Big(\frac{2\pi k}{L}\Big)^2 |\psi_k|^2
\label{HL}
\end{equation}
and the nonlinear part 
\begin{equation}\label{HNL}
H_{\rm NL}(t)=\frac{\sigma}{2L}\int_{0}^{L}|\psi|^4dx.
\end{equation}

 We now assume that the Fourier modes at initial time 
$\psi_{0k}= \psi_{k}(t=0) =|{\psi_{0k}}|e^{i \phi_{0k}}$ 
are  complex random variables.  The complex  field \eqref{eq:psi} is then a  random periodic solution of the 1D-NLSE. No particular
hypothesis about  statistical properties of $\psi_{0k}$ needs to be
introduced  at this step 
but we will show in Sec. \ref{sec:gauss_stat} that the main result  of our analysis can be simplified if the initial statistics
of the random wave field is assumed to be Gaussian. We consider random initial
conditions for which  $N$,  $H_{\rm L}(t=0)$ and $H_{\rm NL}(t=0)$ are all $\mathcal{O}(1)$.
This is typically achieved  by taking the initial power
spectrum $n_{0k}=|\psi_{0k}|^2$ with the characteristic width $\Delta k \simeq 1$, which
implies that the typical spatial size of the initial random fluctuations
is also of
the order of unity and much larger than the internal coherence length (that is $\epsilon$). Such random waves are  often called {\it partially coherent}, particularly in the statistical optics context \cite{Mandel_wolf}.

Given the 1D-NLSE evolution of individual realizations of the random field $\psi(x,t)$  the challenge is to determine the associated evolution of its statistical characteristics  such as the probability density function (PDF) of the amplitude $|\psi|$, the power spectrum $|\psi_k|^2$ etc.
The particular  objective of this paper is to determine the short-time
evolution of the normalized fourth moment $\kappa_4(t)$ 
 defined as
\begin{equation}\label{kappa4}
  \kappa_4(t)=\frac{\langle \frac{1}{L}\int_0^L |\psi(x,t)|^4 \, dx \rangle}
        {\langle  \frac{1}{L}\int_0^L |\psi(x,t)|^2 \, dx \rangle^2},
\end{equation}
where the brackets $\langle \dots \rangle$ denote  ensemble average that
is made over a large number of realizations of the random process $\psi(x,t)$. 
The independence of Fourier modes composing the signal implies statistical homogeneity, hence the average over space  $\lim\limits_{L \to \infty} \frac{1}{L}\int_0^L f(\psi) \, dx$ of an arbitrary function $f(\psi)$ of the random complex field $\psi$ is equivalent to the ensemble
average  over a large number of realizations. Nevertheless, it is convenient for our purposes in this paper
to use the double (space and ensemble) averaging in the definition \eqref{kappa4} of $\kappa_4$. 
We also note that the limit of a large box ($L \to \infty$) 
corresponds to the situation when a large number of random Fourier components
are included inside the core part of the Fourier spectrum, thus
fulfilling the pre-requisites of the central limit theorem for the initial condition.

The fourth  moment \eqref{kappa4} is an important characteristic of the PDF of a random process that quantifies the ``heaviness`` of its tail. In particular,  it can be used to characterize the deviation from Gaussianity  in the course of evolution, when the initial statistics is  Gaussian, in which case   $\kappa_4$ is known to be equal $2$ \cite{Onorato:16,Koussaifi:18}. The determination of $\kappa_4$ is particularly relevant to the rogue wave studies as the formation of a ``heavy tail'' of the PDF is associated with the frequent appearance of large-amplitude events in the random process' realizations \cite{Onorato:13, Walczak:15, Onorato:16, Randoux:16}.

\begin{figure}[h]
\includegraphics[width=9.cm]{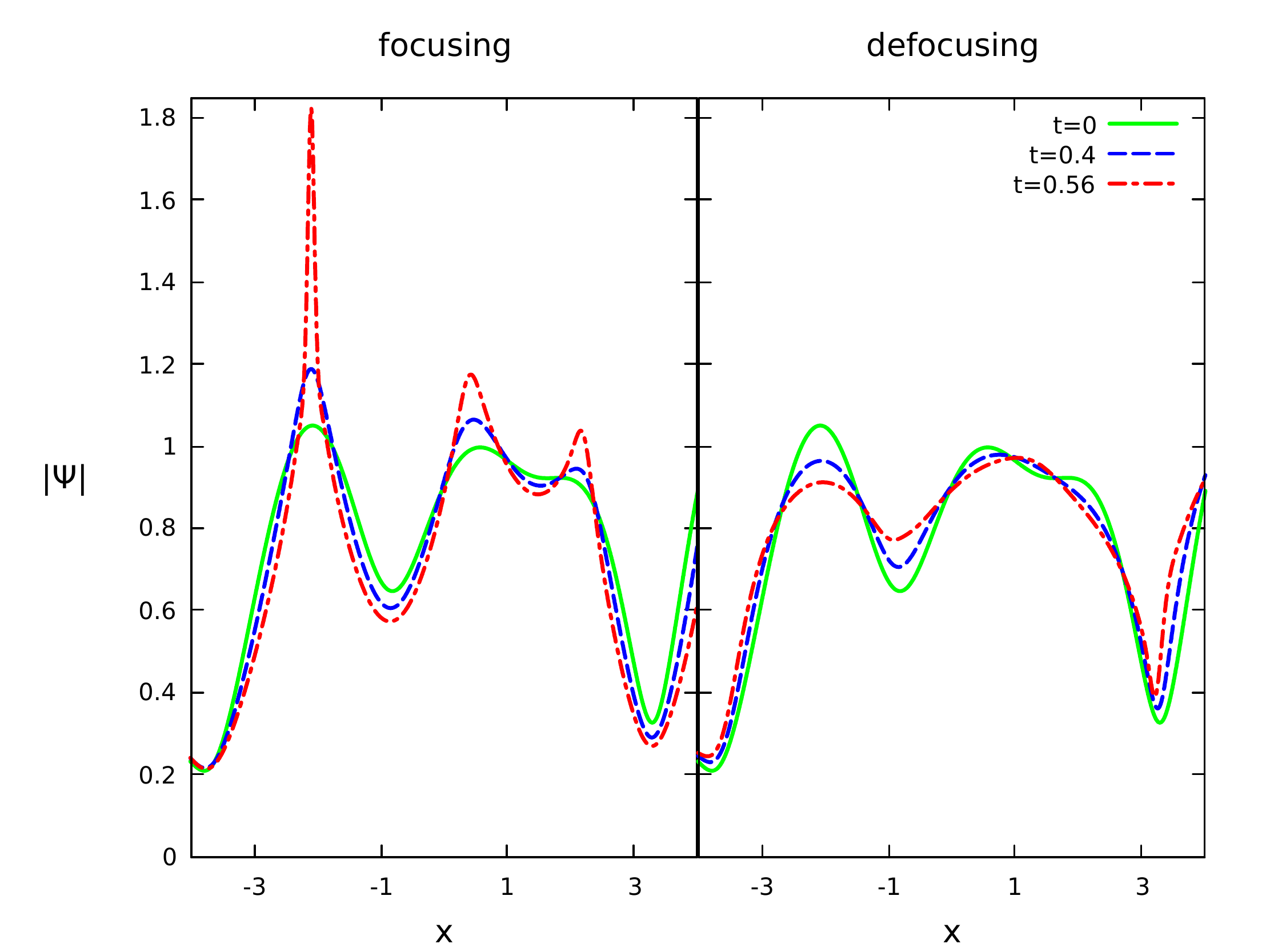}
\caption{Numerical simulations of Eq. (\ref{eq:NLS}) ($\varepsilon=0.1$) showing the time evolution
  of a random field having Gaussian statistics at initial time $t=0$ (green lines).
  (a) Focusing regime ($\sigma = +1$). At short evolution time ($t<0.56$ in the plot)  the self-focusing
  dynamics produces bright peaks
  having the amplitude that grows in time. (b) Defocusing regime ($\sigma = -1$). The
  self-defocusing dynamics induces the decrease in time
  of the amplitudes of  random
  peaks. After some time (not reached in the plot), the random wave
  develops  gradient catastrophes that
  are regularized by dispersive effects leading to the generation of breather structures
  in the focusing regime and of dispersive shock waves in the defocusing regime. 
}
\label{fig:dynamics}
\end{figure}

Fig. \ref{fig:dynamics} shows a typical initial evolution of a random wave in the
regime where the cubic (Kerr) nonlinearity dominates linear dispersive effects, which corresponds to the semiclassical regime described by Eq.~\eqref{eq:NLS} with $\varepsilon \ll 1$ (in the numerics we took $\varepsilon=0.1$).
As shown in Fig. \ref{fig:dynamics}, the self-focusing dynamics tends to produce bright peaks while the self-defocusing
dynamics leads to a decrease of the peak amplitudes but is accompanied 
by steepening of  slopes in the random amplitude profile. While only the short-time
evolution of the wave system is shown in Fig. 1, a longer development leads to the formation of gradient catastrophes -- the explosion of the first derivatives of the wave's profile. These gradient catastrophes  have qualitatively different geometrical nature in the defocusing regime (the wave-breaking singularity \cite{Pomeau:2008}) and the focusing regime (the elliptic umbilic singularity \cite{dubrovin_universality_2008-1}).
 In both cases the gradient catastrophes are regularized by dispersive effects via
the generation of nonlinear short wavelength oscillations: breather structures in the focusing regime \cite{Bertola:13} and 
dispersive shock waves  in the defocusing regime (see \cite{El:2016} and references therein).  For convenience, we shall call the initial nonlinear evolution preceding the formation of  gradient catastrophes, the ``pre-breaking stage'' in both defocusing and focusing regimes. The advantage of the semi-classical, dispersive-hydrodynamic  approach  employed in this paper is that it enables one to asymptotically separate the pre-breaking and post-breaking stages of the evolution, which exhibit qualitatively different behaviors and require very different  analytical methods for their descriptions.

The starting point of our analysis is the evolution of the nonlinear part $H_{\rm NL}$
of the Hamiltonian. Differentiating \eqref{HNL} we obtain 
\begin{equation}
\frac{d H_{\rm NL}}{dt}=\frac{\sigma}{L}\int_0^L |\psi|^2 \, \Big [\psi \frac{\partial \psi^*}{\partial t} + \psi^*\frac{\partial \psi}{\partial t} \Big ] \; dx. \\
\end{equation}
Using Eq. (\ref{eq:NLS}) and integrating by parts, one readily finds  
\begin{equation}
\frac{dH_{\rm NL}}{dt}=\frac{\sigma\varepsilon}{L}\int_{0}^{L}Im[(\psi_x\psi^*)^2]dx.  
\label{eq:dHnldt}
\end{equation}
Now, using the   Madelung transformation
\begin{equation}
\psi=\sqrt{\rho}e^{i\frac{\phi}{\varepsilon}} , \qquad  u=\frac{\partial\phi}{\partial x},
\label{Madelung}
\end{equation} 
Eq. (\ref{eq:dHnldt}) can be rewritten as
\begin{equation}
\frac{dH_{\rm NL}}{dt}=\frac{\sigma}{L}\int_{0}^{L}\rho u\rho_xdx.
\end{equation}


Noticing from \eqref{kappa4}, \eqref{Npart}, \eqref{HNL} that 
\begin{equation}\label{kappa4t}
\kappa_4(t)=\frac{2\langle H_{\rm NL}\rangle}{\sigma \langle N \rangle ^2}
\end{equation}
  one obtains  
\begin{equation}
\frac{d \kappa_{4}}{dt}=\frac{2}{\sigma \langle N \rangle^2}\frac{d\langle H_{\rm NL}\rangle}{dt}=\frac{2}{\langle N \rangle ^2L}\int_{0}^{L}\langle\rho u\rho_x\rangle dx.  
\label{eq:dk4dt}
\end{equation}


We now derive an analytical expression for  $\kappa_4(t)$ for short evolution times, $t\ll1$. If the dispersion parameter is small, $\varepsilon \ll 1$, the initial dynamics are dominated by nonlinearity.
To describe these dynamics analytically, we  consider the semi-classical limit of the 1D-NLSE \eqref{eq:NLS}  which is found by applying the Madelung transform \eqref{Madelung} and letting $\varepsilon \to 0$. Assuming smooth evolution of $\rho(x,t)$ and $u(x,t)$ in the pre-breaking regime,  we obtain in the limit $\varepsilon \to 0$
the following well-known set of nonlinear geometric optics equations
\cite{forest_exact_2009,Wabnitz:13b,Fatome:14,Kodama1995, Randoux:17} \begin{equation}\label{shallow}
\begin{cases}
\rho_t + (\rho u)_x =0, \\ 
 u_t+uu_x - \sigma\rho_x=0.
\end{cases}
\end{equation}
If $\sigma=-1$, equations (\ref{shallow}) are identical to the shallow-water equations
for an incompressible fluid with $\rho>0$ and $u$ interpreted as the fluid depth and
the depth-averaged horizontal fluid velocity, respectively. In the nonlinear fiber optics context,
$\rho$ represents the instantaneous optical power and $u$ represents the instantaneous frequency
(or chirp) \cite{Wetzel:16}. 

Rigorous proofs of the point-wise convergence, as $\varepsilon \to 0$, of solutions of the 1D-NLSE \eqref{eq:NLS} to the solutions of the dispersionless system \eqref{shallow} with the same initial data, prior to the formation of gradient catastrophe, can be found for certain classes of initial data in \cite{jin:1999} (defocussing) and in \cite{Kamvissis, Tovbis:04} (focussing). Some important particular exact solutions of system \eqref{shallow} for the focusing case have been found as early as in 1960-70s  (see \cite{talanov_1965, akhmanov_self-focusing_1966, gurevich_exact_1970}). A detailed mathematical analysis of the pre-breaking dynamics in the defocusing case can be found in  \cite{forest_exact_2009} (see also \cite{Moro:14} for the special case of the wave breaking into vacuum).

It follows from the above consideration that, to study  the pre-breaking dynamics of partially coherent waves in 1-D NLSE \eqref{eq:NLS} we need to be able to describe  {\it random} solutions of system \eqref{shallow} obtained by evolving initial data  $\rho(x,0)$, $u(x,0)$ with given statistics (e.g. corresponding to the Gaussian statistics of $\psi$).  The study of such solutions has been recently initiated   for the defocusing case in the context of the interaction of random Riemann waves in fiber optics \cite{Randoux:17} (see also \cite{Suret:17}). In this connection one must stress that the term ``pre-breaking dynamics'' is understood here in the probabilistic sense, as for random initial data there is always a non-zero probability of having gradient catastrophe at any, arbitrarily small, moment of time. However, due to initial data having typical size $\Delta x = \mathcal{O}(1)$, we assume that for small $\varepsilon$ the contribution of such early gradient catastrophes to the statistics  is negligibly small.

To this end, with the short-time, pre-breaking evolution in mind,  we look for the solutions of Eqs. (\ref{shallow}) in the form of the  time power series
expansions for the realizations $\rho(x,t)$ and $u(x,t)$: 
\begin{equation}\label{eq:series}
\begin{aligned} & \rho(x,t)=\rho_0(x)+\rho_1(x)t+\rho_2(x)t^2+\rho_3(x)t^3+\mathcal{O}(t^4) \\ & u_0(x,t)=u_1(x)t+u_2(x)t^2+u_3(x)t^3+\mathcal{O}(t^4) \, .
\end{aligned}
\end{equation}

 In \eqref{eq:series} we assumed   that initially,  $u_0 =0$, which agrees with  typical physical condition $u_0\ll\rho_0$  satisfied in standard realistic experimental conditions. Indeed timescales of amplitude and phase in partially coherent waves are generally similar ($\mathcal{O}(1)$ here). Considering the normalizations given by the Eq. (\ref{Madelung}), this means that the derivative of the phase $\frac{\partial(\phi_0/\epsilon)}{\partial x}=\mathcal{O}(1)$ and thus $u_0=\frac{\partial\phi_0}{\partial x}=\mathcal{O}(\epsilon)$ whereas $\rho_0 =\mathcal{O}(1)$. This assumption is for example satisfied in the experiments on the propagation of partially coherent light through optical fibers, see \cite{Randoux:17}.\\

Substituting Eqs. (\ref{eq:series}) into Eqs. (\ref{shallow}) we obtain 
\begin{equation}\label{eq:rho_u_time}
\begin{aligned} & \rho(x,t)=\rho_0-\frac{1}{4}\sigma[\rho_0^2]_{xx}t^2+\mathcal{O}(t^4), \\ &u(x,t)=\sigma\rho_{0x} t-(\frac{1}{12}[\rho_0^2]_{xxx}+\frac{1}{3}\rho_{0x}\rho_{0xx}) t^3+\mathcal{O}(t^4).
\end{aligned}
\end{equation}
Next, substituting Eqs.~\eqref{eq:rho_u_time} into Eq. (\ref{eq:dk4dt}) and
integrating in time, we obtain the following expression for the time evolution of the
normalized fourth moment of the field amplitude: 
\begin{equation}
\begin{aligned}
 \kappa_4(t)-\kappa_4(0) = & \frac{\sigma t^2}{\langle N\rangle^2L}\int_{0}^{L}\langle \rho_0\rho_{0x}^2\rangle dx\\ &-\frac{t^4}{2\langle N\rangle^2L}\int_{0}^{L}\langle\frac{2}{3}\rho_0^2\rho_{0x}\rho_{0xxx} \\ & + \frac{17}{6}\rho_0\rho_{0x}^2\rho_{0xx}+\frac{1}{2}\rho_{0x}^4\rangle dx +\mathcal{ O}(t^6).
\end{aligned}
\label{eq:k4_t}
\end{equation}

Eq. (\ref{eq:k4_t}) is our main general result. One can make now two important observations. The first one is that Eq.~\eqref{eq:k4_t} shows that the normalized fourth-order
moment $\kappa_4(t)$ of the field evolves quadratically with time at leading order for $t \ll 1$. The second observation is that Eq. (\ref{eq:k4_t})
 shows that  the increasing or decreasing nature of the time evolution of $\kappa_4(t)$ is determined by the value taken by $\sigma$.
In the focusing regime ($\sigma=+1$), $\kappa_4(t)$ is an increasing function of
time which means that the nonlinear evolution of the wave field is characterized
by PDFs that exhibit heavy tailed deviations from the initial statistical distribution.
On the other hand, in the defocusing regime ($\sigma=-1$) $\kappa_4(t)$ becomes a decreasing function
of time which implies  low-tailed deviations from the initial statistics occurring
in this regime. The statistical features described by Eq. (\ref{eq:k4_t}) are in full qualitative agreement
with the results that have been recently obtained in numerical and experimental investigations of  integrable turbulence \cite{Randoux:16,Randoux:14,Walczak:15,randoux2016livre,Suret:16,Suret:17}.

Let us emphasize that the decreasing or increasing nature of the time evolution of $\kappa_4$ has
also been shown to be determined by the defocusing or focusing nature of the propagation regime
for {\it weakly} nonlinear dispersive random waves that are described by the 1D-NLSE \cite{Janssen:03}.
Theoretical approaches that have been used in the weakly nonlinear regime are based on the
wave turbulence theory and they consist in deriving quasi-kinetic equations for the lowest order moments
of the wave field \cite{Suret:11,Janssen:03}. Dispersion plays crucial role in that consideration. Our work is based on a completely different, dispersive-hydrodynamic
approach, where dispersive effects  are initially not of dominant but of perturbative nature.

\section{Initial conditions with Gaussian statistics}\label{sec:gauss_stat}

Eq. (\ref{eq:k4_t}) represents a general result that is derived with the only assumption
that $\varepsilon \ll 1$. As we already stressed, it is valid before the typical time of the gradient catastrophe occurrence, i.e. for $t \ll 1$
(for random initial conditions with typical scales for $\rho$- and $x$-variations at $\mathcal{O}(1)$).
Importantly, Eq. (\ref{eq:k4_t}) is derived without any assumption on the nature of
the initial statistics of the random wave field. In this section we show that Eq. (\ref{eq:k4_t})
can be further simplified if the random wave field taken as initial condition has  Gaussian
statistics. To this end, we assume that the random initial field $\psi(x,0)$  is composed of a 
linear superposition of a large number of {\it independent} random Fourier
modes $\psi_{k}(t=0)=\psi_{0k}=|{\psi_{0k}}|e^{i \phi_{0k}}$,   so that by the central limit theorem $\psi(x,0)$ is a Gaussian
random field  ~\cite{Nazarenko}.

In the random phase and amplitude (RPA)
model, $|\psi_{0k}|$ and $\phi_{0k}$ are both taken as randomly-distributed 
variables \cite{Nazarenko}. 
Here, we will mainly use the so-called random phase (RP) model 
in which only the phases $\phi_{0k}$ of the Fourier modes are considered as 
being random \cite{Nazarenko}. In this model, the phase of each Fourier 
mode is randomly and uniformly distributed between $-\pi$ and $\pi$. 
Moreover, the phases of separate Fourier modes are assumed to be uncorrelated so that 
$\langle e^{i\phi_{0k}}e^{i\phi_{0k'}}\rangle= \delta_{k}^{k'}$. In the above expression, 
the brackets, as usual,  represent the averaging  over  an ensemble 
of many realizations of the random process; $\delta_{k}^{k'}$ is the 
Kronecker symbol defined by $\delta_{k}^{k'}=1$ if $k=k'$ and $\delta_{k}^{k'}=0$ 
if $k\neq k'$. With the assumptions of the RP model described above, 
the statistics of the initial field is homogeneous, which means that all statistical
moments of the initial complex field $\psi(x,t=0)=\psi_0(x)$ do not depend
on $x$ \cite{Picozzi:07,Picozzi:14}. This RP description of the initial
random field has been shown to describe in a satisfatory way many experiments
performed in the field of integrable turbulence \cite{Randoux:16,Randoux:14,Walczak:15,randoux2016livre,Suret:16,Suret:17,Onorato:04,Onorato:05,Onorato:01,onorato2005modulational}.

Given the delta-correlation of the random phases, the second moment of a field composed from the linear superposition of a large number
of {\it independent} Fourier components having Gaussian statistics is readily evaluated as
\begin{equation}\label{eq:2nd-order}
\big <\psi_{k}\psi_{k'}^*\big > = n_k \, \delta_k^{k'},
\end{equation}

and the  sixth moment can be factored into products of the second moments by using Wick's decomposition~\cite{Nazarenko} 
\begin{equation}\label{eq:4th-order}
\begin{aligned}
  \langle\psi_{k_1}\psi_{k_2}\psi_{k_3}\psi_{k_4}^*\psi_{k_5}^*\psi_{k_6}^* \rangle=n_{k_1}n_{k_2}n_{k_3} && \\
             [\delta_{k_4}^{k_1}\delta_{k_5}^{k_2}\delta_{k_6}^{k_3}+
               \delta_{k_4}^{k_1}\delta_{k_5}^{k_3}\delta_{k_6}^{k_2}+
               \delta_{k_4}^{k_2}\delta_{k_5}^{k_1}\delta_{k_6}^{k_3}+&& \\
               \delta_{k_4}^{k_2}\delta_{k_5}^{k_3}\delta_{k_6}^{k_1}+
               \delta_{k_4}^{k_3}\delta_{k_5}^{k_1}\delta_{k_6}^{k_2}+
               \delta_{k_4}^{k_3}\delta_{k_5}^{k_2}\delta_{k_6}^{k_1}
             ]
\end{aligned}
\end{equation}
Now, using Eq. (\ref{eq:4th-order}), one can evaluate the coefficient for the  $\mathcal{O}(t^2)$ term in the expansion  (\ref{eq:k4_t}):
\begin{equation}\label{eq:t_square_term}
\begin{aligned}
 \frac{\sigma}{\langle N\rangle^2L}&\int_{0}^{L}\langle \rho_0\rho_{0x}^2\rangle dx  = \frac{\sigma}{\langle N\rangle^2}\sum_{k_1,...k_6} \Big(\frac{2 i \pi}{L}\Big)^2 \delta_{k_4+k_5+k_6}^{k_1+k_2+k_3} \\ &\times\langle\psi_{0k_1} \psi_{0k_2}\psi_{0k_3}\psi_{0k_4}^*\psi_{0k_5}^*\psi_{0k_6}^* \rangle (k_2-k_5)(k_3-k_6),
\end{aligned}
\end{equation}
where we have used the notation $\psi_{0k_i} = \psi_{k_i}(0)$ for the Fourier component at $t=0$. Using Eq. (\ref{eq:4th-order}), we obtain the following expression
for the short time evolution of the fourth-order moment of a random wave field that has Gaussian
statistics at initial time, i.e. $\kappa_4(0)=2$:
\begin{equation}\label{eq:k4_time_bis}
\begin{aligned}
\kappa_4(t)-\kappa_4(0)&=-\frac{\sigma}{\langle N\rangle^2}\sum_{k_1,k_2,k_3}n_{0k_1}n_{0k_2}n_{0k_3}\Big(\frac{2 \pi}{L}\Big)^2\\&\times[-2(k_2-k_3)^2]t^2+\mathcal{O}(t^4),
\end{aligned}
\end{equation}
where $n_{0k_i}=n_{k_i}(0)$ are the components of the  power spectrum at $t=0$. Using Eqs. (\ref{Momentum}), (\ref{Npart}), (\ref{HL}) and taking into account that $P=0$
for our random Gaussian  field,
we can finally rewrite Eq. (\ref{eq:k4_time_bis}) as
\begin{equation}\label{eq:k4_time_final_order2}
\kappa_4(t)-\kappa_4(0)=8\sigma \langle H_{\rm L}(0)\rangle t^2+\mathcal{O}(t^4).
\end{equation}
(Note that  $H_{\rm L} = \mathcal{O}(1)$ and $H_{\rm NL} = \mathcal{O}(1)$ whereas the linear part of the Hamiltonian (\ref{eq:H}) is $\mathcal{O}(\epsilon^2)$).

A similar, but somewhat lengthy, calculation permits one to obtain a more accurate expression
that includes $\mathcal{O}(t^4)$ correction (see Appendix \ref{appendixa}):
\begin{equation}
\begin{aligned}
\kappa_4(t)-&\kappa_4(0)=8\sigma \langle H_{\rm L}(0)\rangle t^2+\Bigg[\frac{208}{3} \langle H_{\rm L}(0)\rangle^2\\&+4\langle N\rangle\langle\sum_k\Big(\frac{2 \pi k}{L}\Big)^4|\psi_{0k}|^2\rangle\Bigg]t^4+\mathcal{O}(t^6).
\end{aligned}
\label{KHl}
\end{equation}

Eqs. (\ref{eq:k4_time_final_order2}) and (\ref{KHl}) show that the time evolution
of the fourth moment of the initially Gaussian random  wave field is determined by the linear part 
$ H_{\rm L}(0)$ of the Hamiltonian computed for the initial condition.

Eq. (\ref{KHl}) can be further simplified if we assume that the shape of the Fourier power spectrum
of the initial random field is described by a Gaussian
\begin{equation}\label{eq:CI_gauss}
|\psi_{0k}|^2=n_0e^{-\frac{k^2}{(\Delta k)^2}}.
\end{equation}
The amplitude $n_0 \in \mathbb{R^+}$ has to be determined from the normalization
condition provided by Eq. (\ref{Npart}).
The linear energy density determined from Eq. (\ref{HL})
is $\langle H_{\rm L}\rangle=\frac{N(\Delta k)^2}{4}$, and we can finally rewrite
Eq.~\eqref{KHl} only in terms of the density of particles (or optical power) $N$ 
and of the width $\Delta k$ of the initial Fourier spectrum
\begin{equation}\label{eq:k4_gauss}
\kappa_4(t)-\kappa_4(0)=2\sigma N(\Delta k)^2t^2+\frac{22N^2(\Delta k)^4}{3} t^4 + \mathcal{O}(t^6). 
\end{equation}

\section{Numerical simulations}

In this section, we use numerical simulations of Eq. (\ref{eq:NLS}) to
investigate the range and the degree of validity of the semi-classical 
approach to the statistics of integrable turbulence presented in Secs. \ref{sec:main} and \ref{sec:gauss_stat}.
The initial condition used in our numerical simulations is a random complex field having Gaussian
statistics.  The amplitudes of the Fourier components are taken to be distributed
according to  Eq. (\ref{eq:CI_gauss}). In our numerical
simulations, the spectral phases $\phi_{0k}$ are random, statistically independent real numbers, uniformly 
distributed between $-\pi$ and $+\pi$. The width $\Delta k$ of the
initial spectrum profile \eqref{eq:CI_gauss} is taken to be unity ($\Delta k=1$), and the value
of $\varepsilon$ in \eqref{eq:NLS}  is taken to be $0.1$. The numerical
simulations are performed by using a pseudo-spectral method with 
the numerical box having  size $L=256$ that is discretized by using $2^{16}$ points.
\begin{figure}[h]
\includegraphics[width=9.cm]{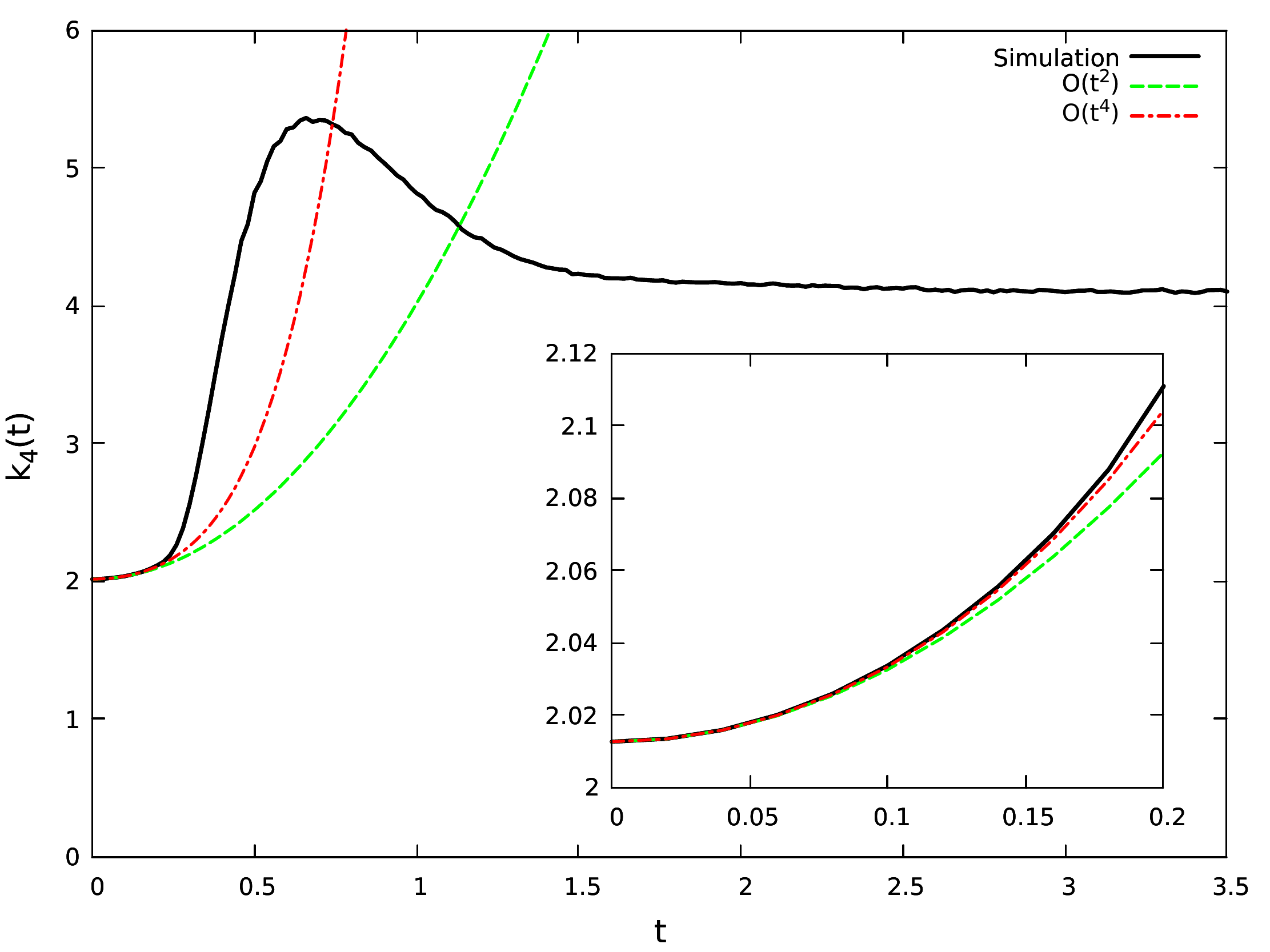}
\caption{ Black line: numerical simulations of
  Eq. (\ref{eq:NLS}) with $\varepsilon=0.1$ in the focusing regime ($\sigma=+1$) showing the time evolution of the normalized fourth-order moment of the random field having at $t=0$ the Gaussian statistics and the Fourier spectrum defined by Eq. (\ref{eq:CI_gauss}) with
 $\Delta k=1$, $N=1$. Green line: analytical  result given by
  Eq. (\ref{eq:k4_gauss}) at leading order in $t^2$. Red line:  Analytical
  result given by Eq. (\ref{eq:k4_gauss}) including $t^2$ and $t^4$ evolution terms.
  The inset shows an enlarged view of the evolution of $\kappa_4(t)$ for $0<t<0.2$.
}
\label{fig:k4_focusing}
\end{figure}

Fig. \ref{fig:k4_focusing} shows the time evolution of the normalized
fourth moment $\kappa_4(t)$ of the random wave field in the focusing
regime ($\sigma=+1$). The curve plotted with
black line represents the result of the numerical simulation of Eq. (\ref{eq:NLS}).
At time $t=0$, the theoretical value taken by $\kappa_4(t)$ is $2$ due to the
Gaussian nature of the initial statistics.  As can be seen from Fig.~\ref{fig:k4_focusing}, in the numerical experiments the value taken by $\kappa_4(t)$ at $t=0$ slightly differs from $2$ (see the inset in Fig. \eqref{fig:k4_focusing}) because the  conditions of the central limit theorem are not perfectly fulfilled in our numerical simulations.  Indeed, because of the finite value of $L$, the number of Fourier modes in the spectrum  given by Eq. \ref{eq:CI_gauss} is finite, in particular, in the Full Width at Half Maximum (FWHM) we count $\frac{\sqrt{2\ln 2} L}{\pi}=95$ modes. Importantly, the deviation of the initial condition from  Gaussian statistics affects only the value of $\kappa_4(0)$ in Eq. \eqref{eq:4th-order} but not the evolution. 
In the initial (before the formation of a gradient catastrophe) stage of the nonlinear evolution of the random wave, $\kappa_4(t)$
is at first an increasing function of time that later reaches a maximum around $t\sim 0.6$.
Then $\kappa_4(t)$ becomes a decaying function of time that reaches a stationary
value around $\sim 4$ at long evolution time. A similar evolution of $\kappa_4(t)$
has already been evidenced in numerical simulations presented in ref. \cite{Onorato:16}.

The curves plotted with green and red lines in Fig. \ref{fig:k4_focusing}
show monotonic evolutions of $\kappa_4(t)$ that are obtained from Eq. (\ref{eq:k4_gauss}). 
In particular the curves plotted in the inset of Fig. \ref{fig:k4_focusing} 
 clearly reveal a very good quantitative agreement
between numerical and theoretical results. In particular, a better
agreement between numerics and theory is obtained by including the
fourth-order correction term found in the
time expansion of the solution, see Eq. (\ref{eq:k4_gauss}).
A significant quantitative disagreement is found
between our theoretical results and the numerical simulation at evolution
times greater than $\sim 0.2$. This arises from the fact that our approach
is only valid at evolution times that are shorter that the typical wave breaking time
(the pre-breaking description). The significant occurrence of wave breakings 
at evolution times greater than $\sim 0.2$ has strong influence on the wave evolution
and subsequently the wave statistics in a way that cannot be accounted for
by using our pre-breaking treatment. 

\begin{figure}[h]
\includegraphics[width=9.cm]{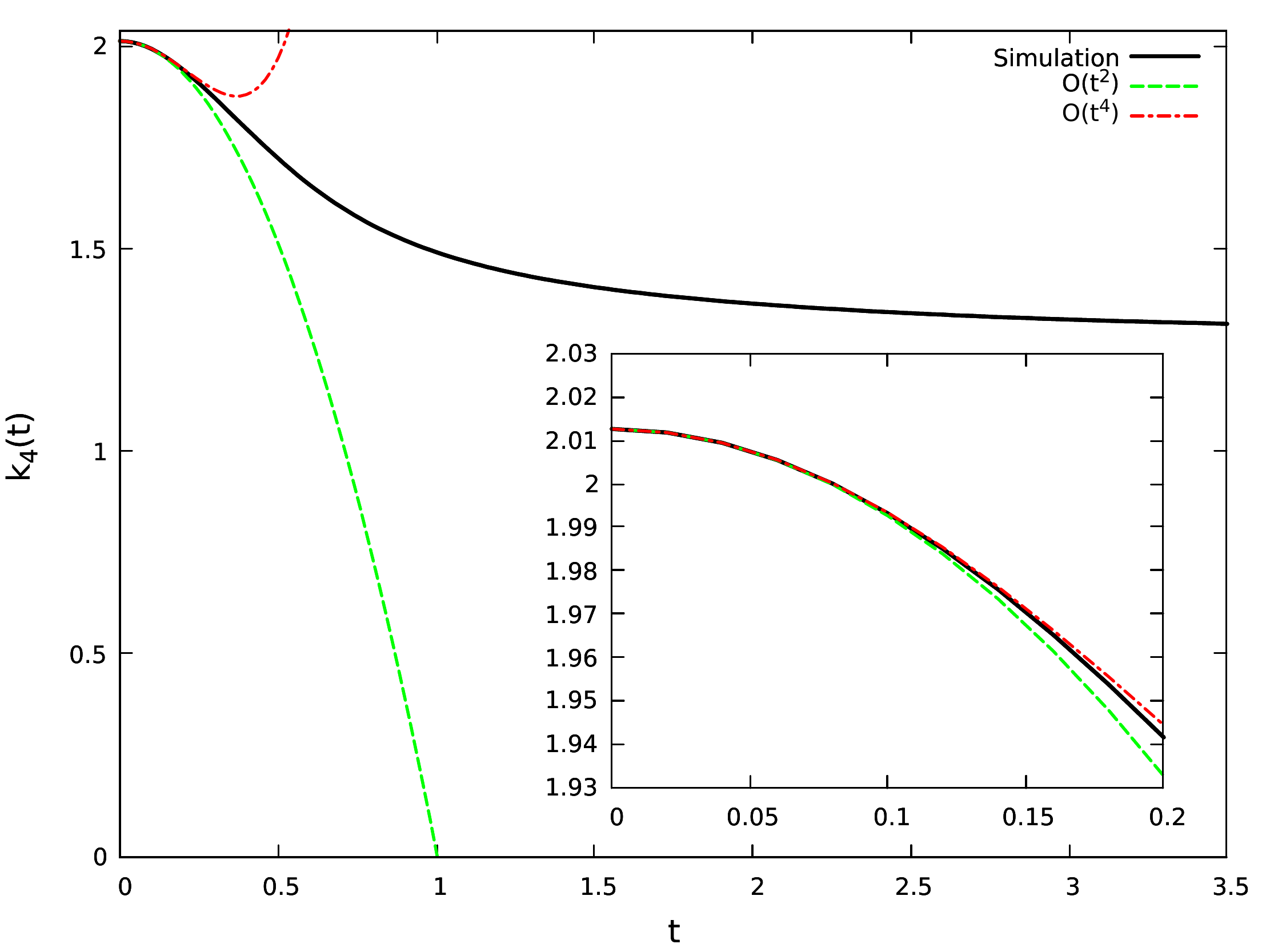}
\caption{Black line: numerical simulations of
  Eq. (\ref{eq:NLS}) with $\varepsilon=0.1$ in the defocusing regime ($\sigma=-1$) showing the time evolution of the normalized fourth-order moment of the random field having at $t=0$ the Gaussian statistics and the Fourier spectrum defined by Eq. (\ref{eq:CI_gauss}) with
 $\Delta k=1$, $N=1$. Green line: analytical  result given by
  Eq. (\ref{eq:k4_gauss}) at leading order in $t^2$. Red line:  Analytical
  result given by Eq. (\ref{eq:k4_gauss}) including $t^2$ and $t^4$ evolution terms.
  The inset shows an enlarged view  of $\kappa_4(t)$ plot for $0<t<0.2$.
}
\label{fig:k4_defocusing}
\end{figure}

Fig. \ref{fig:k4_defocusing} shows the comparison between the numerical simulation
of Eq. (\ref{eq:NLS}) and the theoretical result given by Eq. (\ref{eq:k4_gauss})
in the defocusing regime ($\sigma=-1$). In the defocusing regime, $\kappa_4(t)$ is
a monotonically decreasing function of time, as already evidenced
in ref. \cite{Onorato:16}. As for the focusing regime, a very good quantitative
agreement is obtained between numerics and the theory at short evolution time ($t<0.2$),
i.e. before the typical occurrence of gradient catastrophes. 

\section{Conclusion}
In this paper we  have undertaken an analytical study of the problem of the  evolution of a random wave field in the 1D-NLSE for both focusing and defocusing regimes. This has been done from the perspective
of dispersive hydrodynamics, a semiclassical theory of
nonlinear dispersive waves exhibiting two distinct spatio-temporal scales:
the long scale specified by initial conditions and the short scale by the internal coherence length
(i.e., the typical size $\varepsilon$ of the coherent structures) \cite{Biondini:16}. 
This scale separation enabled us to split the time evolution of the nonlinear
random wave system (integrable turbulence) into the initial,  ``pre-breaking'' stage, preceding the formation of  gradient catastrophes, when the evolution of the 1D NLSE wave field is almost everywhere smooth, and the ``post-breaking'' stage characterised by the generation of short-scale nonlinear oscillations (breathers or dispersive shock waves depending on the focusing vs. defocusing character of the 1D-NLSE).

Our work is concerned with the initial, pre-breaking stage of the semi-classical integrable turbulence, when the dynamical and
statistical features can be analytically described  in terms  of
random solutions of the dispersionless (nonlinear geometric optics) system \eqref{shallow}. As a result, we have derived a simple  asymptotic formula describing the 
evolution of the normalized fourth moment of the random wave field. This formula, applied to the problem of the 1D-NLSE evolution of random field having initial Gaussian statistics, describes  the initial stage of the formation of heavy tails of the PDF of the field amplitude in the focusing case and the formation of low tails in the defocusing case.

    Recently, an exact and general identity that relates the changes in the statistical properties of the wave field to the changes of its Fourier spectrum has been derived by using the Hamiltonian structure of 1D-NLSE~\cite{Onorato:16}. In other words, the knowledge of the fourth order moment also provides the description of spectral properties. The general description of the stationary state of integrable turbulence and the theoretical prediction of the fourth order moment is still an open fundamental question.    In the weakly nonlinear regime, the wave turbulence approach provides  a statistical description of the nonlinear propagation of random wave fields in 1DNLSE systems~\cite{Janssen:03,Mori:06,Soh:10,Suret:11}.  Recently, using an approach based on the so-called large deviation theory, it has been  shown that rogue waves obey a large deviation principle, i.e. the heavy tails of the PDF of the random wave field are  dominated by single realizations \cite{Dematteis:18, Dematteis:18:NLS}. This  approach is very promising but  does not provide a simple formula for the evolution of the statistics. In this article we have demonstrated that the semi-classical approach is an extremely powerful tool enabling one to describe in a simple way the early stage of integrable turbulence {\it in the strongly nonlinear/ small dispersion} regime.  The proposed methodology  can be applied to the description of partially coherent random nonlinear waves described by other integrable equations, including shallow water waves described by the KdV equation and its extensions. In particular, the pre-breaking statistics of bi-directional random shallow water waves is equivalent to that described by the defocusing 1D-NLSE and studied in this paper.

The semi-classical approach to the statistics of random waves in integrable systems is general and can be used beyond the short-time asymptotic regime. It is known very well that, in the semi-classical limit the evolution of nonlinear dispersive waves after the gradient catastrophe point is described by the so-called Whitham modulation equations \cite{whitham74} governing the behaviour of the averaged integrals of motion, and replacing the dispersionless system \eqref{shallow} (see \cite{kamchatnov_nonlinear_2000, El:2016, el_dam_2016, tovbis_semiclassical_2016}  and references therein for the application of the Whitham theory to the defocusing and focusing 1D-NLSE).   Such an extension of the proposed method to longer times  is very promising but also highly nontrivial.


\section{Appendix: Computation of $\mathcal{O}(t^4)$ corrections for the case of Gaussian statistics at $t=0$} \label{appendixa} 

Here we provide the simplified expressions for the three terms that are found in the
integral giving the coefficient of the $O(t^4)$ term in Eq. (\ref{eq:k4_t}).
To obtain these expressions, we assume Gaussian statistics at the initial time and
use Eq. (\ref{eq:2nd-order}) and Eq. (\ref{eq:4th-order}) to obtain

\begin{widetext}

\begin{equation}
  \begin{aligned}
    -\frac{1}{3\langle N\rangle^2L} \int_{0}^{L}\langle\rho_0^2\rho_{0x}\rho_{0xxx}\rangle dx &= -\frac{1}{3\langle N\rangle^2}\Big(\frac{2 i \pi}{L}\Big)^4 \times\sum_{k_1,...k_6} \langle\psi_{0k_1}\psi_{0k_2}\psi_{0k_3}\psi_{0k_4}\psi_{0k_5}^*\psi_{0k_6}^*\psi_{0k_7}^*\psi_{0k_8}^* \rangle       \\ \times &\delta_{k_5+k_6+k_7+k_8}^{k_1+k_2+k_3+k_4}(k_3-k_7)(k_4-k_8)^3 =48 \langle H_{\rm L}\rangle^2+4\langle N\rangle\langle\sum_k\Big(\frac{2 \pi k}{L}\Big)^4|\psi_{0k}|^2\rangle
\end{aligned}
\end{equation}
\begin{equation}
\begin{aligned}
  -\frac{17}{12\langle N\rangle^2L}\int_{0}^{L}\langle \rho_0\rho_{0x}^2\rho_{0xx}\rangle dx &= -\frac{17}{12\langle N\rangle^2}\Big(\frac{2 i \pi}{L}\Big)^4 \times\sum_{k_1,...k_6}\langle\psi_{0k_1}\psi_{0k_2}\psi_{0k_3}\psi_{0k_4}\psi_{0k_5}^*\psi_{0k_6}^*\psi_{0k_7}^*\psi_{0k_8}^* \rangle\delta_{k_5+k_6+k_7+k_8}^{k_1+k_2+k_3+k_4}  \\
  \times &(k_2-k_6)(k_3-k_7)(k_4-k_8)^2= \frac{136}{3} \langle H_{\rm L}\rangle^2
\end{aligned}
\end{equation}
\begin{equation}
\begin{aligned}
-\frac{1}{4\langle N\rangle ^2L}\int_{0}^{L}\langle \rho_{0x}^4\rangle dx &=-\frac{1}{4\langle N\rangle^2} \Big(\frac{2 i \pi}{L}\Big)^4  \times \sum_{k_1,...k_6}\langle\psi_{0k_1}\psi_{0k_2}\psi_{0k_3}\psi_{0k_4}\psi_{0k_5}^*\psi_{0k_6}^*\psi_{0k_7}^*\psi_{0k_8}^* \rangle \delta_{k_5+k_6+k_7+k_8}^{k_1+k_2+k_3+k_4}   \\ \times& (k_1-k_5)(k_2-k_6)(k_3-k_7)(k_4-k_8)=-24 \langle H_{\rm L}\rangle^2
\end{aligned}
\end{equation}

\end{widetext}

\begin{acknowledgments}
 This work  was  supported by EPSRC grant
EP/R00515X/1 (GE) and Dstl grant DSTLX-1000116851(GR, GE, SR). It has also been partially supported by the Agence Nationale de la Recherche through the LABEX CEMPI project (ANR-11-LABX-0007)   and by the Ministry of Higher Education and Research, Hauts-De-France Regional Council and European Regional Development Fund (ERDF) through the Contrat de
Projets Etat-R\'egion (CPER Photonics for Society P4S). GE and GR  thank the PhLAM laboratory at the University of Lille for hospitality and partial financial support (GE). 
\end{acknowledgments}


%

\end{document}